# WHERE DO I RANK? AM I HAPPY?: LEARNING INCOME POSITION AND SUBJECTIVE-WELLBEING IN AN INTERNET EXPERIMENT

EIJI YAMAMURA

*Seinan Gakuin University, Japan*

A tailor-made internet survey experiment provides individuals with information on their income positions to examine their effects on subjective well-being. In the first survey, respondents were asked about their household income and subjective well-being. Based on the data collected, three different respondents' income positions within the residential locality, within a group of the same educational background, and cohort were obtained. In the follow-up survey for the treatment group, respondents are informed of their income positions and then asked for subjective well-being.

Key findings are that, after obtaining information, a higher individual's income position improves their subjective well-being. The effects varied according to individual characteristics and proxies. Prosocial respondents do not report higher SWB even after they know that their income position is higher. The effects of income position are not observed for respondents aged less than 45 years when life satisfaction is used as a proxy for SWB.

# 1. INTRODUCTION

'I think; therefore, I am.' This is the foundation for human existence. However, people compare themselves with the relevant others, which is correlated with their subjective well-being ( SWB)[1]. Many studies assess whether one's and others' income influences SWB (e.g., Clark and Oswald, 1996; Senik, 2004; Ferrer-i-Carbonell, 2005; Luttmer, 2005; Latif, 2016; Burida and ).

Recent studies showed income rank is better than the average income of the reference group to consider the relative income effect (Clark et al., 2009a; Boyce et al., 2010; Brown et al., 2008; Burkhouser et al., 2016; Budria and Ferrer-i-Carbonell, 2019). However, most existing studies concerning SWB are based on an implicit assumption that people know others' income or their relative income position in society. In the analysis of redistribution preference closely related to the SWB analysis[2], researchers doubt the assumption, so they conduct a survey experiment. They examine the impact of biased perceptions about individuals' income position on their preferences (Cruces et al., 2013;

---

[1] The seminal work of Easterlin (1974) asked whether a rise in the income of all improves the happiness of all. The question was trigger for many works to consider social comparison and adaptation to income in social science (e.g., Alesina et al., 2004; Ferrer-I-Carbonell and Frijters, 2004; D'Ambrosio and, Frick, 2012; Clark and Georgellis, 2013).
[2] Several studies examine how income comparison influences SWB and redistribution preference (Clark and Senik, 2010; Senik, 2009).

Karadja et al., 2017)[3]. They found that their preference changed by being informed of their real income position because their perceived income position was biased[4]. This study applies the experimental approach employed in these studies to investigate how and the extent to which provision of one's real income position influences respondents' SWB. I used Internet experiments that have the advantage of gathering representative samples of society and larger sample sizes compared to laboratory experiments[5].

Through survey experiments, researchers assessed how income position influences preference only after providing the information (Cruces et al., 2013; Karadja et al., 2017). Hence, they did not consider preferences before learning their income positions. It is more appropriate to consider how information changes subjective values by comparing before and after given information (Shigeoka and Yamada, 2019). Furthermore, it is important to set reference groups to measure the income position. Many studies have analysed how the SWB of individuals depends on the reference group (e.g., Clark and Senik, 2010; Senik, 2009; Brown et al., 2015; Goerke and Pannenberg, 2015; Noy and Sin, 2021; Clark et al., 2021). These studies collected subjective information of subjective intensity regarding comparison and intention to compare to whom. However, they do not know the income of others to whom they intend to compare themselves. For instance, there is a possibility that people are more likely to overestimate others' income, as they are more likely to compare. Further, people who feel happier may underestimate others' income and perceive their income ranks higher. Therefore, it is appropriate to exogenously provide several relative income positions and then examine their influence to mitigate these biases.

---

[3] The effect of self-perceived relative income on SWB is examined in existing studies (Yu, 2019). However, its finding is considered to suffer from bias.
[4] After learning true information, people's view concerning various issues changes. For instance, for people who have negative views on immigration, providing information about immigrants leads them to support immigration if their misperceptions about the immigrants' characteristics are corrected (Grigorieff et al., 2020).
[5] Internet experiments were increasingly used in economics (e.g., Ifcher et al., 2020; Kuziemko et al., 2015; Shigeoka and Yamada, 2021; Yamada and Sato, 2013).

This study deals with endogenous bias and misperceptions. The contribution of this study is the first to compare the same respondent's SWB before and after being informed of several income positions measured by various reference groups[6]. Further, following the study to investigate the effect of the provision of income position on redistribution preference (Karadja et al., 2017; Budria and Ferrer-i-Carbonell, 2019), this study also assesses the process and extent to which the effect of learning the correct income position depends on an individual's characteristics[7].

In the first survey, I gathered respondents' SWB (happiness level and life satisfaction), positive reciprocity, generalised trust, household income, and basic information about sex, educational background, and age. Two weeks later, the second follow-up survey was distributed randomly to two-fifth of the respondents who were informed about their true income positions, and others were not informed. All respondents were asked about the SWB again.

---

[6] Norwegian citizens are allowed to observe the income of everyone since 2001. Based on the natural experiment, Perez-Truglia (2020) provide the evidence higher transparency increased the gap in happiness between richer and poorer citizens.
[7] Research using survey data found that the relation between income position and SWB differs according to individual characteristics (Burdria and Ferrer-i-Carbonell, 2019).

Moreover, there are some proxy variables for SWB. Among them, happiness and life satisfaction are widely used and are interpreted similarly (e.g., Clark et al., 2008; Clark and Senik, 2010). However, the effects of income position on SWB are larger when happiness, rather than life satisfaction, is used as a proxy for SWB (Perez-Truglia, 2020). Some researchers emphasise the difference between happiness and life satisfaction (Chui et al., 2016; Diener et al., 2004; 2009). On the one hand, happiness can be regarded as a 'pleasant emotion', an affective, short-lived reaction tied to specific events. Another definition is 'the frequent experience of positive emotions over time' (Lyubomirsky et al., 2005, p. 806). On the other hand, life satisfaction refers to one's ongoing evaluation of the conditions of life as a whole, which presumably requires cognitive processing. According to Diener et al. (2004, p. 205), a person 'can be satisfied with one's life, and yet experience little pleasant affect and vice versa'. Income comparison is thought to have a different influence on happiness and life satisfaction because of this difference. Hence, as proxies for SWB, both happiness and life satisfaction were used to examine the effect of providing real income positions.

Consistent with existing studies, respondents who know that they are in a higher income position feel happier and more satisfied with their lives. However, regardless of the real income position, learning the information reduces respondents' SWB. Prosocial respondents do not report higher SWB even after knowing their income position is higher, whereas others report higher SWB if they are informed of a higher-ranked income position. Additionally, the positive effect of income position on happiness level is observed in most cases, even after dividing the sample into sub-samples. Contrastingly, the positive effect of the position on life satisfaction disappears for females, low-educated, and older people, although the effect is observed for males, educated, and young people.

The remainder of this paper is organised as follows. Section 2 describes the experimental strategy and provides an overview of the data. Section 3 provides an econometric framework and proposes testable hypotheses. Section 4 explains the estimation approach, and Section 5 provides the estimation results and their interpretation. The final section offers some conclusions and implications for future research.

## 2. EXPERIMENTAL DESIGN OF INTERNET SURVEY AND DATA

A flowchart of the simple experiment is shown in Figure 1. As explained below, the same respondents participated in the first and follow-up surveys to answer the same questions about SWB. The difference is that they do not know their position in the reference group, but are informed of the position directly before answering the SWB.

<insert figure 1 around here>

2.1. *The first survey*

In Japan, the Nikkei Research Company has experience in academic research on internet surveys (Clark et al., 2021). Therefore, they were commissioned to conduct a nationally representative web survey covering all parts of Japan on 25-30th October 2018[8].

A total of 9,130 participants participated in the survey. In the first survey, 7855 observations were gathered, which was reduced to 5,647 in the follow-up survey. Eventually, the response rate reached 62%. In the first survey, as basic information, respondents were asked about residential locality, chosen from 47 prefectures, ages, educational background[9], and annual household income from the previous year[10].

---

[8] Web users are possibly different from non-web users. However, an official survey on information technology indicates that in 2015 nearly 100% of Japanese people in the 20–29, 30–39, and 40–49 age groups are web users. Even for older age groups, the percentage of web-users is over 90% for people aged 50–59 and 80% for people aged 60–69. Therefore, the sampling method through the Internet is unlikely to suffer bias.

[9] There are eight choices categorized by final educational background; (1) primary school, (2) Junior high school, (3) High school, (4) vocational school, (5) college, (6) University, (7) Graduate school, (8) do not want to answer

[10] There are 13 choices (in 10 thousand Japanese yen, which is approximately 100 US$); (1) below 100, (2) 100-199, (3) 200-399, (4) 400-599, (5) 600-799, (6) 800-999, (7) 1,000-1,199, (8) 1,200-1,399, (9) 1,400-1,599, (10) 1,600-1,799, (11) 1,800-1,999 (12) over 2,000, (13) do not want to

They were then asked to report their state of subjective view about trust and the SWB. As proxies for prosocial characteristics, respondents were asked about reciprocity and trust, and there are five choices.

*Do you agree with the view 'Most people can be trusted'?*

*Do you agree with the view 'If someone does me a favour, I am prepared to return it'*

*1 (Strongly disagree) to 5 (Strongly agree).*

Concerning SWB, there are two questions and five choices:

*Question about happiness 'How would you rate your current level of happiness?'*

*1 (very unhappy) to 5 (very happy).*

*Question about happiness 'To what extent are you satisfied in your life?'*

*1 (very dissatisfied) to 5 (very satisfied).*

2.2. *The follow-up survey*

Two weeks after the first survey, the second follow-up survey was distributed to the respondents who have completed the questionnaire in the first survey. The two-period panel data is constructed through the first and follow-up surveys. A random experiment

---

answer. In this study, the mid-point of each range is used as household income. In the case of (12), the upper bound is arbitrarily defined to be 2,300. However, the rate of this case is only 1%, so estimation is hardly biased by the definition of upper bound value.

is conducted using the information gathered from the first survey. Subjects are randomly divided into three groups. Around 7,855 subjects were invited and 5,647 observations were gathered. These are divided into 2,331 in the treatment group, 2,170 in the control group (A), and 1,126 in the control group (B).

In the treatment group, two-fifths of the subjects were informed of their true income position in happiness and life satisfaction. Therefore, they learn their income position and answer the SWB.

An individual's SWB depends not only on one's income but also on the income level of neighbouring people (e.g., Clark and Oswald, 1996; Luttmer, 2005; Card et al. 2012)[11]. However, various groups can be used as reference groups. According to previous studies (e.g., Senik 2009; Clark and Senik 2010; Goerke and Pannenberg, 2015; Clark et al 2021), it is crucial to choose the reference group because the effects of the income comparison differ possibly according to the reference group. However, according to the 'aspirational income hypothesis' proposed by Khalil et al. (2021), individuals aspire toward status and therefore prefer to identify with a group more successful than they are[12]. If this holds, endogenous bias occurs if the reference group is determined by an individual's preference.

---

[11] Preference preferences are is negatively correlated with the average incomes of individuals' ethnic and religious groups (Quattrociocchi 2018).
[12] Falk and Knell (2004) argued that people tend to compare themselves to similar others.

To avoid the bias, in this study, three reference groups were set based on the data gathered by the first survey: (1) income position within a residential locality (prefecture), (2) the position in the same cohort[13], and (3) the position within the same educational background[14].

As illustrated by the arrow of the dashed line in Figure 1, before answering the questions, different individuals' income ranks are informed as percentages of income rank.

*'Based on the information as below, please answer the questions'*

*Using household income data from the first survey, your income position is calculated as follows:*

*In the group which consists of those whose educational background is the same as you, your income rank is the top 15 %.*

*In the group consisting of those in the same age range as you, your income rank is the top 18 %.*

*In the group of those who resided in the same prefecture as you, your income rank is the top 11 %.'*

---

[13] Cohort groups are defined by every ten birth years.
[14] In the questionnaire, there are seven categories of educational backgrounds. However, in some of the groups, only a few subjects are included. For calculating the individual income position, educational backgrounds are roughly divided into three groups; High educated groups who have graduated from university or graduate school where total schooling years are 16 or more. Intermediate educated groups who have graduated from vocational schools or college, where total schooling years are 14. Low educated groups have graduated from high school, Junior high school, or primary school, where total schooling years are 12 or less.

The questionnaire explained that the larger the percentage, the higher the number of raked respondents. Percentages slightly varied according to reference groups.

There are two types of control groups. To consider the relative income effect on the SWB, neighbouring countries can serve as a reference group (Becchetti et al., 2013). As the control group A, two-fifths of the subjects were informed of Japan's rank of income inequality by comparing three countries (Germany, USA, and China) arbitrarily selected. Information about each country's rank is based on the Gini coefficient obtained from the World Bank data[15]. Japan's rank is used as a placebo because it does not provide any information on an individual's rank. However, in contrast to the first survey, information is provided directly before answering the SWB. As for Control group B, one-fifth of the subjects were not provided with any information and so are required to answer the SWB, which is just the same way in the first survey.

For Control group A, all respondents see the same information as follows because there is no variation in national rank even for different subjects.

*'Based on the information below, please answer the questions'*

*Compared to other countries, Japan's rank of income inequality is as follows (the higher*

---

[15] World Bank Gini rank. https://www.indexmundi.com/facts/indicators/SI.POV.GINI/rankings (accessed on Oct 19, 2018).

*rank, the smaller the inequality).*

   *27th Germany     37th Japan    100th USA   110th China*

*2.3. Data*

As explained in Subsection 2.2., the number of respondents included in the estimation was 2,331 for the treatment group, 2,170 for Control group A, and 1,126 for Control group B. They respond to the first and follow-up survey. Therefore sub-sample sizes used in the estimations are 4,662, 4,340, and 2,252 for the treatment group, the Control group A, and Control group B, respectively.

For the balance check, Table 1 presents the results of the mean-difference test in the first wave. There was no significant difference in the variables between the treatment and control groups. The exceptional cases are 'RANK_AGE' between the treatment and Control group B, but its statistical difference is only at the 10 % level. Overall, respondents have almost the same characteristics as the groups. Therefore, the respondents of the treatment and control groups can be comparable.

In addition to Table 1, to verify the experiment's validity, Figures 2-3 compare the distribution of SWB between the treatment and control groups. Figure 2 compares the happiness levels between groups in the first survey. Respondents are likely to choose '3' or '4' in all groups, and its distributions are almost the same between groups. Figure 3 reports the distribution of life satisfaction and demonstrates similar results as Figure 2. Overall, there is no difference in SWB before informing the correct information of income position for the treatment group.

<figure 2 around here>

<figure 3 around here>

In Figure 4, the kernel density is used to show that the distribution of household income is skewed toward the left, and there is no difference between the groups. Considering Table 1 and Figures 2-4, the validity of the experimental analysis was verified using the online survey.

<Figure 4 around here>

## 3. THE ECONOMETRIC MODEL AND HYPOTHESES

to examine the effect of providing the correct information of income position on the SWB, fixed effects model with interaction terms were conducted. The specification of the estimation model is

HAPPY (or SATISF)$_{it}$ = $\alpha_1$ SECOND$_t$ *RANK_LOCAL$_i$ (LANK_AGE$_i$, or LANK_EDU$_i$)+ $\alpha_2$ SECOND$_t$ + m$_i$ + u$_{it}$,

The dependent variable is the proxy variable, HAPPY (or SATISF)$_{it}$. Suffix '$i'$ is the individual $i$'s SWB at the timing of survey $t$. SECOND is the follow-up survey dummy. RANK_LOCAL is the ranked position in the income distribution of residential locality for the individual $i$, which does not change in the first and second surveys. SECOND *RANK_LOCAL is the key independent variable, which is the interaction term between SECOND and RANK_LOCAL. Instead, as an alternative specification, RANK_AGE$_i$ (or RANK_EDU$_i$) is used to interact with SECOND. m$_i$ is the individual's time-invariant fixed effects. $u_{it}$ is error term.

SECOND *RANK_LOCAL is included to examine the effect of providing the correct information of income position in the locality. In the treatment group, following previous studies (e.g., Brown et al., 2008; Clark et al., 2009a; Powdthervee, 2009; Budria and Ferrer-i-Carbonell, 2019), SECOND *BIAS_LOCAL is expected to have a positive sign in case higher-ranked position increases the SWB.

In the control groups, respondents are not informed of their real income positions, even in the follow-up survey. However, in the Control A group, respondents are newly informed of Japan's rank of income inequality in comparison to Germany, the USA, and China. Hence, the effect of placebo information can be reflected in the sign of SECOND coefficient. Conversely, in Control B, no information is provided, and thus, no variables are predicted to be statistically significant.

Three different reference groups are used to consider how one's income position differs according to the reference group. Further, the effect of income position varies according to the characteristics of the respondents (Budria and Farrer-i-Carbonell, 2019). As widely observed, the SWB is positively associated with prosocial characteristics such as trust (e.g., Helliwell, 2003; 2006; 2007; 2010) and positive reciprocity (Dohmen et al., 2009). Reciprocal people tend to 'see and share others' unhappiness. To the extent that positive reciprocal individuals are empathic and derive disutility from seeing others suffering, we expect that they obtain no satisfaction from occupying a higher rank' (Budria and Farrer-i-Carbonell, 2019, p.341). Similarly, people who tend to trust others have empathy and are unlikely to benefit from their higher-income position. Therefore, *Hypothesis 1* is proposed:

*Hypothesis 1: The SWB of prosocial people is less likely to be influenced by their income position.*

SWB depends on internal reference points, such as the individual's income aspirations for the future (Clark et al., 2008). Brunello (2020) provides evidence that individuals who have completed a vocational high school are more likely to report a higher level of happiness than individuals who have completed an academic degree. One reason is that vocational high school graduates tend to have more moderate aspirations than graduates of academic education (Brunello, 2020). To put it differently, in comparison with less educated people, highly educated people have higher aspirations. It is plausible that high aspiration gives highly educated people an incentive to earn higher income in the future and to earn more than others[16]. In other words, highly educated people intend to compare to relevant others. Here, *Hypothesis 2* is postulated.

*Hypothesis 2: The SWB of highly educated people is more likely to be influenced by their income position.*

Tsurumi et al. (2019) found that the adverse effects of income of the reference group become smaller when respondents are older. This is partly because 'During retirement, people may be less likely to compare themselves with others because they have lost this familiar reference group' (Tsurumi et al., 2019, p.170). Even before retirement, through

---

[16] If individuals' aspiration is higher, they would prefer to join groups with a higher rank than their own. On this assumption, people are less satisfied if others' income is lower (Kalil et al., 2021). Meanwhile, inevitably, endogenous bias occurs. However, the results of this study suffer from bias because three different reference groups are given exogenously.

various experiences in life, people may think more of the various facets of quality of life than material wealth or income position. It seems plausible that older adults are more likely to think of trust and social relations. A higher-income position increases one's sense of economic superiority. However, it results in the coldness of human relations and erodes trust with others. Reduction in social capital cancels out benefits from higher-income positions if the importance of social capital is sufficiently large. Therefore, *Hypothesis 3* was derived:

*Hypothesis 3: The SWB of middle-aged or aged persons is less likely to be influenced by their income position.*

To test this hypothesis, sub-samples were used to examine how the effect of the provision of an individual's income position on the SWB differs according to prosocial and other groups, young and old-middle aged groups, and high and low educational background groups.

## 4. ESTIMATION RESULTS

Tables 2-11 report the results of estimation about SWB using the fixed effects model. Tables 2 and 3 exhibits the baseline results for HAPPY and SATISF, respectively. In Tables 4-11, sub-sample estimation results to compare the effect of learning income position between groups with different characteristics.

4.1. *Baseline Estimation*

In Table 2, the results for the treatment group in columns (1)–(3) show that the coefficient of the cross terms shows a positive sign and its statistical significance. The absolute value of the coefficient of SECOND*RANK_LOCAL is 0.175, which can be interpreted as suggesting that learning household income in the top rank increases happiness level from the first to the follow-up surveys by a 0.175 point on a 5 point scale. These values are slightly larger than 0.143 and 0.156 for SECOND*RANK_AGE and SECOND*RANK_EDU, respectively. Meanwhile, for Control groups A and B, no cross-terms show statistical significance. Therefore, those not informed of their income position do not change their happiness level from the first to the follow-up survey. Considering this together reveals that respondents do not have correct information about their income position beforehand, so the provision of income rank changes their happiness level.

Interestingly, the coefficient of SECOND is a negative sign, and statistical significance at the 1 % level in columns (1)–(6) for the treatment and control group A. Its absolute values are around 1.00, which is considered sizable. However, SECOND is not statistically significant for Control group B. Information about Japan's rank of income inequality is provided for Control group A but not for Control group B. Therefore, happiness level declines if respondents are provided with information related to an individual's income rank or national rank of income inequality. According to Clark and Senik (2010), comparing oneself with others makes people unhappy. However, they examined subjective views about the importance of comparing with others. They noted that 'unhappy people may compare to others more to explain and justify their lower well-being …we are not able to conclude on any causality in this cross-section data (Clark and Senik 2010, p.587)'. Panel data constructed by Internet experiments enables me to control for biases. The results of SECOND strongly support their argument that comparisons are associated with lower levels of happiness.

Table 3 shows that the cross terms yield a significant positive sign in columns (1)-(3), whereas the cross-terms are not statistically significant in columns (4)-(9). Therefore, respondents in the higher-ranked income position are more satisfied with their life only after learning his real income position. They do not know their income position beforehand. Furthermore, for the Treatment group and Control Group A, the significant negative sign was observed. These are similar results to Table 2 for estimation of happiness. However, the absolute values of coefficients shown in Table 3 are smaller than those in Table 2. Therefore, the effects of learning income position on life satisfaction are smaller than on happiness.

4.2. *Comparison between prosocial respondents and others*

4.2.1. *Prosocial people vs others*.

As presented in Table 1, the mean values of reciprocity are slightly greater than four on a five-point scale, which indicates that most respondents chose four or five. Therefore, the group of reciprocal persons is defined as those who chose the highest values of five. Accordingly, reciprocal persons occupy one-third of the whole sample. In Panel A, the cross-terms show a positive sign for all columns. However, statistical significance was only observed for 'Not reciprocal' individuals. Further, its coefficients are approximately two times larger in 'Not reciprocal' individuals than 'Reciprocal' individuals. Cross terms in Panel B do not indicate statistical significance in any column. This is because respondents are not informed of the correct income position. Results of Table 5 are similar to those of Table 4, although coefficients of cross-terms in Table 5 Panel A of the 'Not reciprocal' group are smaller by approximately 0.03 than in Table 4. The impact of income position on life satisfaction is smaller than that on happiness.

The combined results of Tables 4 and 5 are consistent with *Hypothesis 1*.

4.2.2. *People with high generalized trust vs others.*

In Table 1, the mean values of trust are approximately 3.15, on a five-point scale. The sample can be almost divided even if respondents who chose 4 or 5 are defined as the group with high trust, otherwise another group. Panel A of Table 6 shows the positive sign of the cross-terms in all the columns. The 'Not trust' results are statistically significant at the 1% level in all estimations, whereas those of 'Trust' are not significant, except for column (1). Further, the values of coefficients for the 'No Trust' group are around 0.22, which is about two times larger than those of the 'Trust' group. Panel B of Table 6 shows hardly any statistically significant cross-terms, except for column (5).

The results in Table 7 are almost the same as those in Table 6, showing that respondents who are unlikely to trust others are more likely to be satisfied when their income position is higher. However, this tendency is not observed for those who trust others.

Based on the results of Tables 4-7, respondents who tend to trust others and reciprocal respondents do not improve their SWB even if they learn that their true income ranks are high. These results strongly support *Hypothesis 1*.

*4.3. Comparison between high and low educated individuals*

As presented in Table 8, the sample is divided into those who are educated from the university and others. Here, those who are educated from university or more are defined as highly educated people, while others are less educated people. Panels A and B show the results of the Treatment and Control group A, respectively. Panel A shows the significant positive sign of the cross-terms for the fifth and sixth. Highly educated respondents show a higher statistical significance than less-educated ones. Further, the absolute values of the coefficient for highly educated respondents are approximately 0.30, which is three times larger than approximately 0.1 for less educated ones. This implies that the happiness of highly educated people is more influenced by knowing their position than the less educated people. Panel B does not suggest any significant cross-term results, implying that respondents of Control Group A are not influenced by their correct income position if they are not informed. These results are consistent with the results presented in Table 2.

Turning to Table 9, Panel A shows that the cross-terms exhibit a positive sign in all columns. For the highly educated respondents, the cross-terms are statistically significant. The absolute values of the coefficients range from 0.25 to 0.35, almost the same as those in Table 8. Meanwhile, interestingly, no cross-terms are statistically significant, and their coefficients are smaller than 0.09 for the less educated respondents in columns (4)–(6) of Panel A. There would be no difference in perceived income position before and after being informed of their income position if the less educated people's perception about their income position is not biased in the first wave. However, as observed in Table 8, this does not hold because the low educated people feel happier in the follow-up survey than in the first survey if they know the higher ranked income position.

There are possibly various interpretations about the insignificant effect of the cross-terms of the less educated respondents in the treatment group. Students who intend to earn more than others in the future are more motivated to obtain a higher academic background. Contrastingly, those who do not place importance on income position are less likely to study hard to obtain a higher academic background.

The combined results of Tables 8 and 9 are congruent to *Hypothesis 2*.

4.4. *Comparison between those over 45 years and those below 45*.

Table 10 presents the results of the sub-sample divided by ages. The middle-aged group is over 45 years, while the younger group is equal to or less than 45. In this way, the sample size of the middle-aged group was almost equal to that of the young group. In Panel A, a significant positive sign is observed for the cross-terms in all columns. Hence, the higher-ranked income position leads respondents to be happier regardless of their age. The absolute values of its coefficients for young respondents are larger by approximately 0.02 than those of the older ones. Hence, the effects of income position information are slightly larger for young respondents than for old ones. Panel B does not show the statistical significance of the cross-terms, except for columns (4) and (5).

In Table 11, the cross-terms show a positive sign in all columns of Panels A and B. Most of the results are similar to the estimation in Table 10. However, a striking difference was observed in respondents over 45 years of age in the treatment group. Neither cross-terms nor SECOND was statistically significant. Older people's life satisfaction does not change even after they are informed of their correct income position, which is different from the results of the happiness estimation in Panel A of Table 10. Considering Tables 10 and 11 together reveals that higher-income positions increase happiness but not life satisfaction when respondents are equal or over 45 years old. Therefore, *Hypothesis 3* was strongly supported.

4.3. *Discussion*

Budria and Ferrer-i-Carbonell (2019) found that the SWB of prosocial people is unlikely to increase compared to others, even if they are in higher-ranked income positions. However, it is unknown whether respondents know their correct position in their study, although unobserved individual time-invariant fixed characteristics are controlled by using the panel data. Their findings are biased if the respondent's perception of their position differs from the correct position. This study used an experiment to correct the bias and made it evident that the SWB of prosocial persons does not increase even after learning that their income is high. This implies that prosocial persons place much importance on social relations to others and think more of others.

The less educated people feel happier but are not satisfied if they know their income position is higher. Inconsistent results can be explained by differences in happiness and life satisfaction. Happiness can be considered a positive emotion, such as momentary pleasure or joy in reacting to a specific event (Chui et al., 2016; Diener et al., 2004; 2009). According to classical works (Vebren, 1899; Frank, 1985), the type of SWB increases with positional material and immaterial goods (conspicuous consumption and leisure) respectively. However, the findings in this study can be interpreted to mean that life satisfaction is more likely to be rooted in life. Brunello (2020) argued that those who graduated from vocational schools tend to live in small towns where social life is more rewarding and have a less privileged parental background. Naturally, they have more moderate aspirations and are happier than those who graduated from university. The findings of this study are consistent with those of Brunello (2020). Respondents who did not graduate from university are less likely to evaluate their lives by comparing their income with others. It is widely acknowledged that highly educated people can earn higher incomes than less educated ones. People who place more importance on economic success are more likely to take a degree from a university. Accordingly, highly educated people are more satisfied with higher-ranked income positions. Meanwhile, less educated people have a different view of life and philosophy, depending on life satisfaction. Hence, their income position does not influence life satisfaction.

As is widely acknowledged, social capital, such as trust with others, improves SWB (e.g., Helliwell, 2003; 2006; 2007) and community attachment (Binder and Freytag, 2013; Tsurumi et al., 2019). The positive impact of social capital is larger than that of income (Bjørnskov, 2003). Workers are more satisfied with their lives from non-financial job characteristics, such as workplace trust (Helliwell and Huang, 2010). Conspicuous consumption may provoke envy, which has a detrimental effect on personal relationships with surrounding people. Similarly, a relatively higher income position arouses the envy of lower-income people[17]. There is a possibility that conflict occurs between high-and low-income groups[18]. This leads high-income people to prefer income redistribution from rich to poor people (Yamamura, 2016). Particularly, high-income earners residing in areas with larger social capital are more likely to prefer redistribution (Yamamura, 2012). They have been motivated to avoid reducing social capital and SWB. Tsurumi et al. (2019) found that relative income has a smaller effect on SWB among older people. The positive influence of household income on happiness decreases for older adults (Hsieh, 2011). Hence, as one becomes older, people are likely to pursue the meaning of life than conspicuous consumption or relative income position from a long-term perspective. If this is true, social capital formed through intimate relationships with others may become more important than when they were younger.

---

[17] There are possibly both envy and signalling effects in income comparison, which depends on the definition of the reference group (Ifcher et al., 2018) and institutional background. Signalling effects are observed in Senik (2004), which found a positive correlation between the mean income of the reference group and happiness for Russia during the 1994–2000 transition period. A positive

## 5. CONCLUSION

---

correlation is also found in Eastern European Countries (Caporale et al., 2009). Job satisfaction is positively correlated with coworkers used as the reference group (Clark et al., 2009b; Javdani and Krauth, 2020). The correlation possibly changed or differs according to the institutional setting. Income of neighbours reduces SWB in West Germany but increases SWB in East Germany (Knies 2012). In Germany, various institutional reforms were carried out over the period 1991- 2009. Before the reforms, signalling was the dominant concern in East Germany, whereas envy was dominant in Western Germany. However, after the reforms, this dominance was not observed (Welsch and Kuhling 2015).

[18] After the fall of the Berlin Wall, income comparison with West German's people reduces SWB of East German's people. Then, East Germans express a negative attitude towards foreigners (Hyll and Schneider 2018).

Many existing studies have found that individuals' income position is correlated with their SWB. However, various studies have made it evident that perceived income position is biased, and thus correct information on income position should be provided to examine the effect of income position (Cruces et al., 2013; Karadja et al., 2017). This study uses an Internet survey experiment to examine how one's income position influences the SWB and the extent to which the effect of income position differs according to individual characteristics such as prosocial view, educational background, and age.

In a tailor-made internet survey experiment, the first survey gathered respondents' household income and SWB. Based on the data collected, the respondents' income positions are calculated. In the follow-up survey, respondents were informed of their income positions and then answered the SWB.

Key findings are that, after obtaining information, a higher individual's income position improves their subjective well-being. This is consistent with the analysis of redistribution preference which indicates that misperceived income position leads to biased outcomes (Cruces et al., 2013; Karadja et al., 2017). Regardless of the real income position, learning the information reduces respondents' SWB. This can be interpreted as suggesting that people tend to feel irksome when compared to others. Prosocial respondents do not report higher SWB even after they know that their income position is higher. The effects of income position drastically reduce or are not observed for prosocial, low educated, and middle-aged respondents, especially when life satisfaction is used as a proxy. This implies that they place more importance on non-monetary values such as personal relations, the meaning of life, and reason to live from long-term and wide-range perspectives. Human beings are socially existent. This study naturally leads us to reconsider the question about 'I think; therefore, I am' from the viewpoint of relative income and social capital.

REFERENCES

Alesina, A., De Tella, R. and MacCulloch, R., 'Inequality and happiness: are Europeans and Americans different?,' *Journal of Public Economics*, 88, 2009–2042, 2004.

Becchetti, L., Castriota, S., Corrado, L. and Ricca, E.G., 'Beyond the Joneses: Inter-country income comparison and Happiness,' *Journal of Socio-Economics*, 45, 187–195, 2013.

Benjamin, Daniel, J., Hefetz, Ori., Kimball, Miles, S., and Szembrot, Nichole., 'Beyond

happiness and satisfaction: Toward well-being indices based on stated preference,' *American Economic Review*, 104, 9, 2698–2735, 2014.

Binder, M., and Freytag, A., 'Volunteering, Subjective Well–Being and Public Policy,' *Journal of Economic Psychology*, 34, 97–119, 2013.

Bjørnskov, C., 'The Happy Few: Cross–Country Evidence on Social Capital and Life Satisfaction,' *Kyklos*, 56, 3–16, 2003.

Boyce, C.G., Brown, G., and Moore, S., 'Money and Happiness: Rank of Income, not Income, Affects Life Satisfaction,' *Psychological Science*, 21, 471–475, 2010.

Brown, G.D. A., Gardner, J., Oswald, A., and Qian, J., 'Does wage rank affect employees' well-being?,' *Industrial Relations,* 47, 355–389, 2008.

Brown, S., Gray, D., and Roberts, J., 'The relative income hypothesis: A comparison of methods,' *Economics Letters*, 130, 47–50, 2015.

Brunello, Giorgio, 'Happier with Vocational Education?,' IZA Discussion Papers 13739, 2020.

Budria, Santi, and Ferrer-I-Carbonell, Ada., 'Life satisfaction, income comparisons and individual traits,' *Review of Income and Wealth*, 65, 2, 337–357, 2019.

Caporale, G.M., Georgellis, Y., and Yin, Y.P., 'Income and happiness across Europe: Do reference values matter?,' *Journal of Economic Psychology*, 30, 42–51, 2009.

Card, David., Mas, Alexander., Moretti, Enrico., and Saez, Emmanuel., 'Inequality at work: The effect of peer salaries on job satisfaction,' *American Economic Review*, 102, 2981–3003, 2012.

Chui, W.,H., Mathew Y., and Wong, H, 'Gender Differences in Happiness and Life

Satisfaction Among Adolescents in Hong Kong: Relationships and Self-Concept,' *Social Indicators Research*, 125, 1035–1051, 2016.

Clark, A.E, Frijters, P., and Shields, M, 'Relative Income, Happiness, and Utility: An Explanation for the Easterlin Paradox and Other Puzzles,' *Journal of Economic Literature*, 46, 1, 95–144, 2008.

Clark, A., and Georgellis, Y, 'Back to Baseline in Britain: Adaptation in the BHPS,' *Economica*, 80, 496–512, 2013.

Clark, A., Kristensen, N., and Westergaard-Nielsen, 'Economic Stratification and Income Rank in Small Neighborhoods,' *Journal of European Economic Association*, 7, 519–527, 2009a.

Clark, A., Kristensen, N., and Westergaard-Nielsen, 'Job Satisfaction and Co-Worker Wages: Status or Signal?,' *Economic Journal*, 119, 4430–447, 2009b.

Clark, A., and Oswald, A., 'Satisfaction and comparison income,' *Journal of Public Economics*, 61, 359–381, 1996.

Clark, A. E., and Senik, C. 'Who Compares to Whom? The Anatomy of Income Comparisons in Europe,' *Economic Journal*, 120, 573–594, 2010.

Clark, A. E., Senik, C., and Yamada, K., 'The Joneses in Japan: income comparisons and financial satisfaction,' *Forthcoming in Japanese Economic Review*, 2021.

Cruces, G., Perez-Truglia, R., and Tetaz, M., 'Biased perception of income distribution and preferences for redistribution: Evidence from a survey experiment,' *Journal of Public Economics*, 98, 100–112, 2013.

D'Ambrosio, C., and Frick, J, 'Individual Well-being in a Dynamic Perspective,' *Economica*, 79, 284–302, 2012.

Diener, E., Oishi, S., and Lucas, R. E., 'Subjective well-being: The science of happiness and life satisfaction,' In C. R. Snyder and S. J. Lopez (Eds.), The handbook of positive psychology (2nd ed., pp. 187–194). New York: Oxford University Press, 2009.

Diener, E., Scollon, C. N., and Lucas, R. E., 'The evolving concept of subjective well-being: The multifaceted nature of happiness,' In P. T. Costa & I. C. Siegler (Eds.), Recent advances in psychology and aging (pp. 188–219). *Amsterdam: Elsevier*, 2004.

Dohmen, T., Falk, A., Huffman, D., and Sunde, U., 'Homo Reciprocans: Survey Evidence on Behavioural Outcomes,' *Economic Journal*, 119, 536, 592–612, 2009.

Easterlin, R., 'Does Economic Growth Improve the Human Lot? Some Empirical Evidence,' In (David, R., and Reder, R, des), Nations and Households in Economic Growth: Essyas in Honor of Moses Abramovitz, 89-125. New York:

Academic Press, 1974.

Falk, A., and Knell, M., 'Choosing the Joneses: endogenous goals and reference standards,' *Scandinavian Journal of Economics,* 106, 417–435, 2004.

Ferrer-I-Carbonell, Ada., Frijters, P., 'How important is methodology for the estimates of the determinants of happiness?,' *Economic Journal*, 114, 641–659, 2004.

Ferrer-I-Carbonell, Ada., 'Income and well-being: An empirical analysis of the comparison income effect,' *Journal of Public Economics*, 89, 997–1019, 2005.

Frank, R.H., 'Choosing the Right Pond,' Oxford: Oxford University Press, 1985.

Goerke, Laszlo., and Pannenberg, Markus., 'Direct Evidence for Income Comparisons and Subjective Well-Being across Reference Groups,' *Economics Letters*, 137, 95–101, 2015.

Grigorieff, Alex., Roth, Christopher., and Ubfal, Diego., 'Does Information Change Attitudes toward Immigrants?,' *Demography* 57, 1117–1143, 2020.

Helliwell, J. F., 'How's life? Combining individual and national variables to explain subjective well-being,' *Economic Modelling*, 20, 2, 331–360, 2003.

Helliwell, J.F., 'Well-Being, Social Capital and Public Policy: What's New?,' *Economic Journal, Royal Economic Society*, 116, 510, 34–45, 2006.

Helliwell, J.F., 'Well-Being and Social Capital: Does Suicide Pose a Puzzle?,' *Social Indicators Research: An International and Interdisciplinary Journal for Quality-of-Life Measurement*, 81, 3, 455–496, 2007.

Helliwell, J.F., and Huang, H., 'How′s the Job? Well-being and Social Capital in the
34

Workplace,' *Industrial and Labour Relations Review*, 63, 2, 355–389, 2010.

Hsieh, C. M., 'Money and Happiness: Does Age Make a Difference?,' *Ageing and Society*, 31, 1289–1306, 2011.

Hyll, W., and Schneider, L., 'Income comparisons and attitude towards foreigners-evidence from a natural experiment,' *Journal of Comparative Economics*, 46, 2, 634–655, 2018.

Ifcher, John., Zarghamee, Homa., Graham, Carol., 'Local Neighbors as Positives, Regional Neighbors as Negatives: Competing Channels in the Relationship between Others' Income, Health, and Happiness,' *Journal of Health Economics,* 57, 263–276, 2018.

Ifcher, John., Zarghamee, Homa., Houser, Dan., and Diaz, Lina., 'The Relative Income Effect: An Experiment. Experimental Economics,' 23, 4, 1205–1234, 2020.

Javdani, M., and Krauth, B., 'Job satisfaction and co‐worker pay in Canadian firms,' *Canadian Journal of Economics,* 53, 1, 212–248, 2020.

Karadja, M., Mollerstrom, J., and Seim, D., 'Richer (and holier) than thou? The effect of relative income improvement on demand for redistribution,' *Review of Economics and Statistics*, 99, 2, 201–212, 2017.

Khalil, E.L., Aimone, J.A., Houser, D., Wang, S., Martinez, D., and Qian, K., 'The aspirational income hypothesis: On the limits of the relative income hypothesis,' *Journal of Economic Behaviour and Organization*, 182, 229–247, 2021.

Knies, Gundi., 'Income Comparisons among Neighbours and Satisfaction in East and West Germany, Social Indicators Research,' 106, 3, 471–489, 2012.

Kuziemko, I., Norton, M., Saez, E., and Stantcheva, S., 'How elastic are preferences for redistribution? Evidence from randomized survey experiments,' *American Economic Review*, 105, 4, 1478–1508, 2015.

Latif, Ehsan., 'Happiness and Comparison Income: Evidence from Canada,' *Social Indicators Research*, 128, 1, 161–177, 2016.

Luttmer, P. E., 'Neighbors as negatives: Relative earnings and well-being,' *Quarterly Journal of Economics*, 120, 3, 963–1002, 2005.

Lyubomirsky, S., King, L., & Diener, E., 'The benefits of frequent positive affect: Does happiness lead to success?,' *Psychological Bulletin*, 131, 6, 803–855, 2005.

Noy, S., Sin, I., 'The effects of neighbourhood and workplace income comparisons on subjective well-being,' *Journal of Economic Behaviour and Organization*, 185, 918–945, 2021.

Perez-Truglia, Ricardo., 'The Effects of Income Transparency on Well-Being: Evidence from a Natural Experiment,' *American Economic Review*, 110, 4, 1019–1054,

2020.

Powdthavee, N., 'How Important is Rank to Individual Perception of Economic Standing? A Within-community analysis,' *Journal of Economic Inequality*, 7, 225–248, 2009.

Quattrociocchi, Jeff., 'Group income and individual preferences for redistribution,' *Canadian Journal of Economics,* 51, 4, 1386–1418, 2018.

Senik, Claudia., 'When information dominates comparison: Lesson from Russian subjective panel data,' *Journal of Public Economics,* 88, 2099–2123, 2004.

Senik, Claudia., 'Direct Evidence on Income Comparisons and Their Welfare Effects,' *Journal of Economic Behaviour and Organization*, 72, 1, 408–444, 2009.

Shigeoka, Hitoshi, and Yamada, Katsunori., 'Income-Comparison Attitudes in the United States and the United Kingdom: Evidence from Discrete-Choice Experiments,' *Journal of Economic Behaviour and Organization*, 164, 414–438, 2019.

Tsurumi, Tetsuya., Imauji, Atsushi., and Managi, Sunsuke., 'Relative income, community attachment and subjective well-being: Evidence from Japan,' *Kyklos*, 72, 1, 152–182, 2019.

Veblen, T., 'The Theory of the Leisure Class. London [Originally published by Macmillan New York], George Allen and Unwin,' (1899,1949).

Welsch, Heinz., and Kuhling, Jan., 'Income Comparison, Income Formation, and Subjective Well-Being: New Evidence on Envy versus Signalling,' *Journal of Behavioural and Experimental Economics*, 59, 21–31, 2015.

Yamada, Katsunori., and Sato, Masayuki., 'Another Avenue for Anatomy of Income Comparisons: Evidence from Hypothetical Choice Experiments,' *Journal of Economic Behaviour and Organization*, 89, 35–57, 2013.

Yamamura, E., 'Social capital, household income, and preferences for income redistribution,' *European Journal of Political Economy*, 28, 4, 498–511, 2012.

Yamamura, E., 'Social conflict and redistributive preferences among the rich and the poor: testing the hypothesis of Acemoglu and Robinson,' *Journal of Applied Economics*, 19, 1, 41–64, 2016.

Yu, Han., 'The Impact of Self-Perceived Relative Income on Life Satisfaction: Evidence from British Panel Data,' *Southern Economic Journal,* 86, 2, 726–745, 2019.

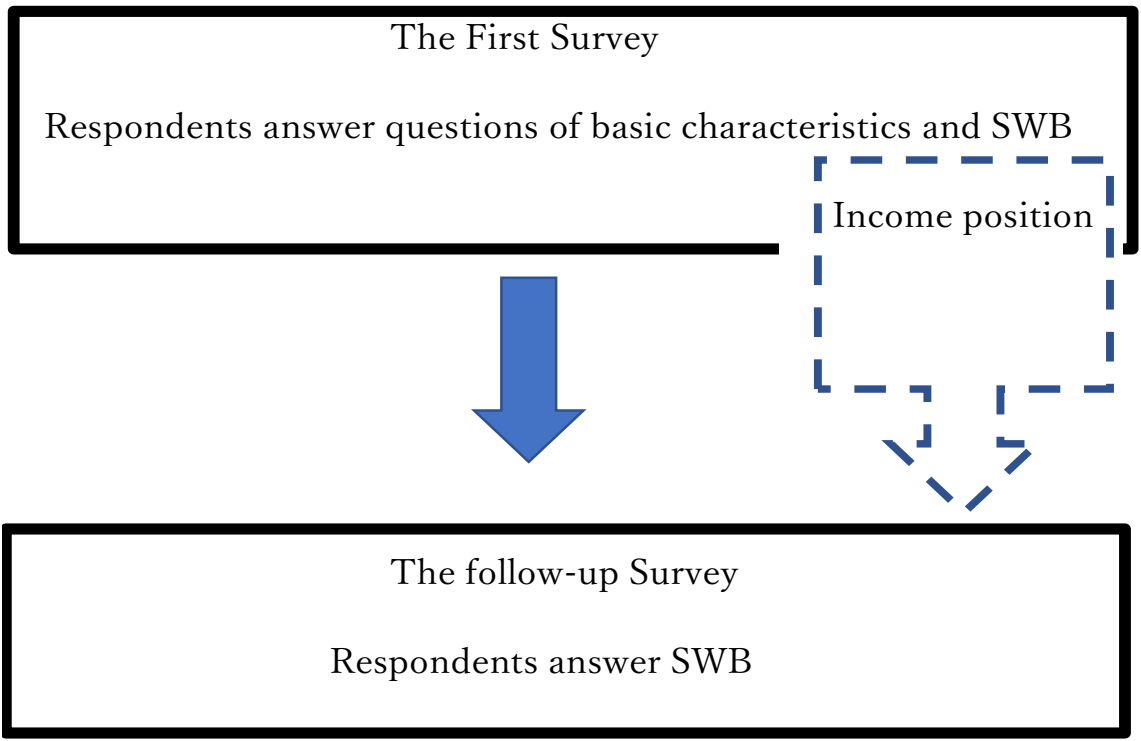

Fig1. Flow-chart of the internet experiment

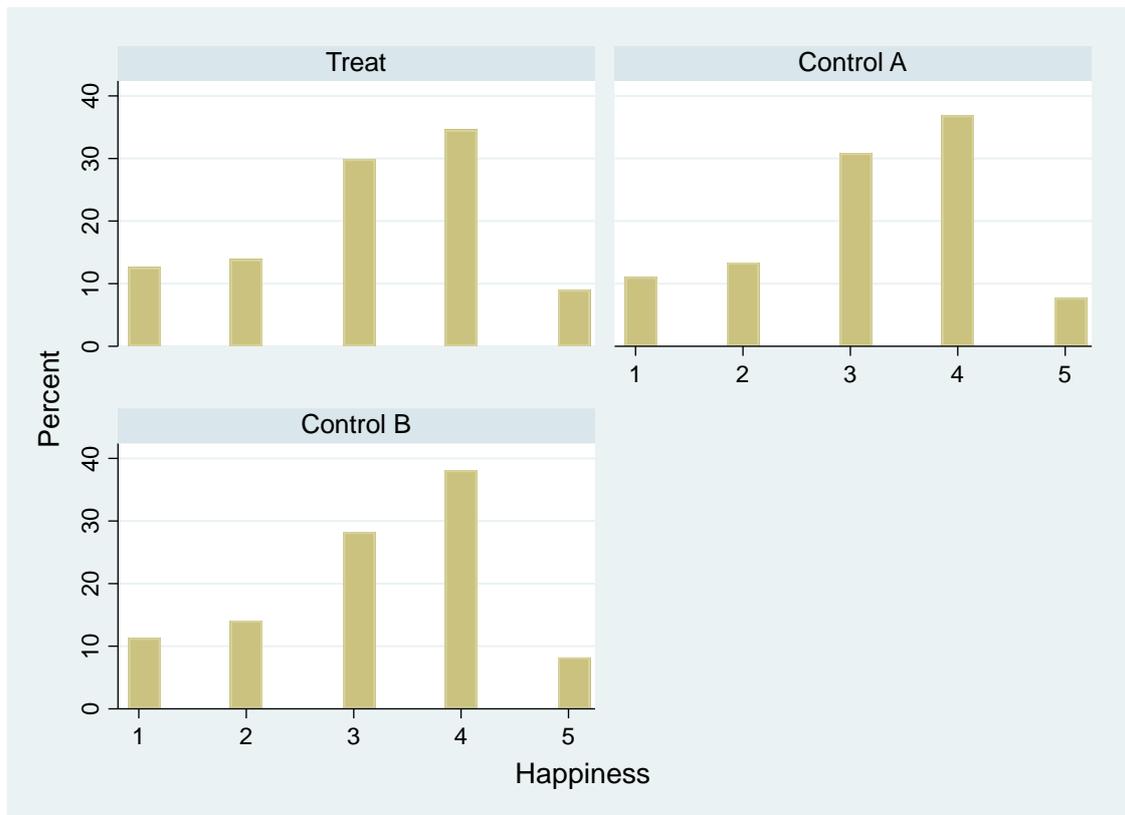

Fig2. Histogram of happiness in the first survey (before correct information about income position).

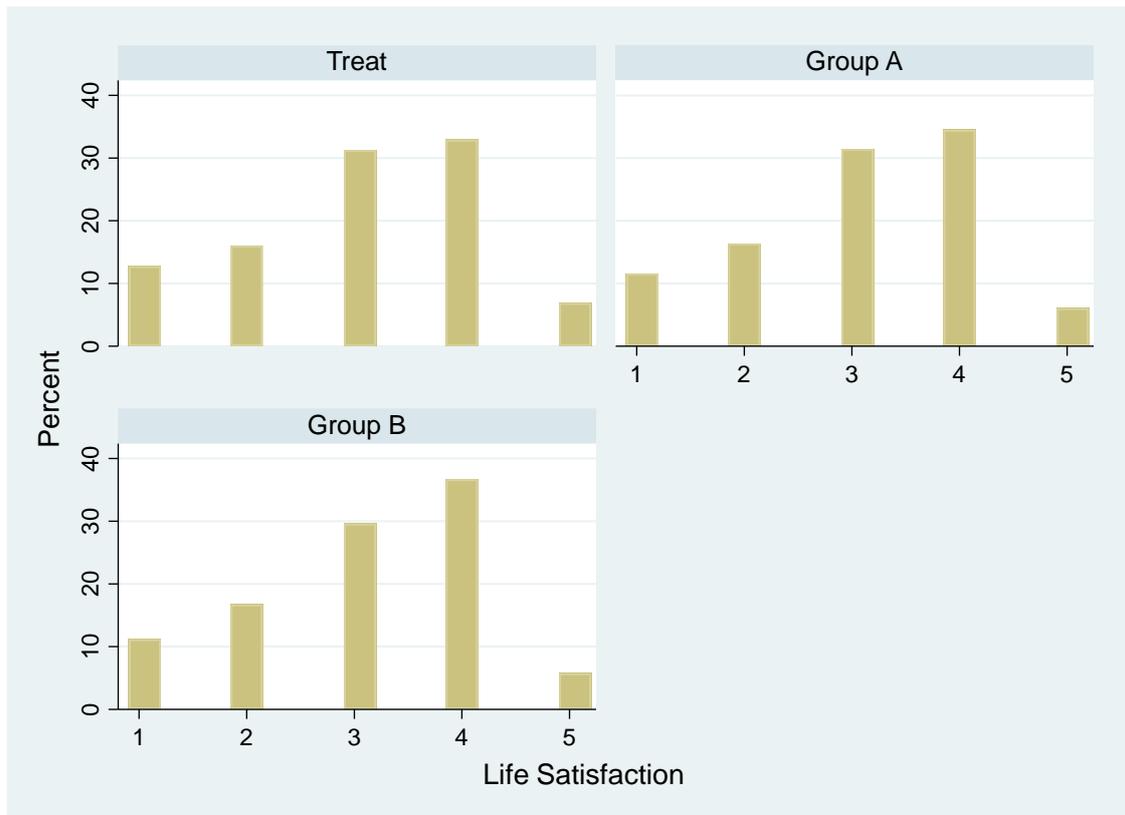

Fig3. Histogram of life satisfaction in the first survey (before correct information about income position).

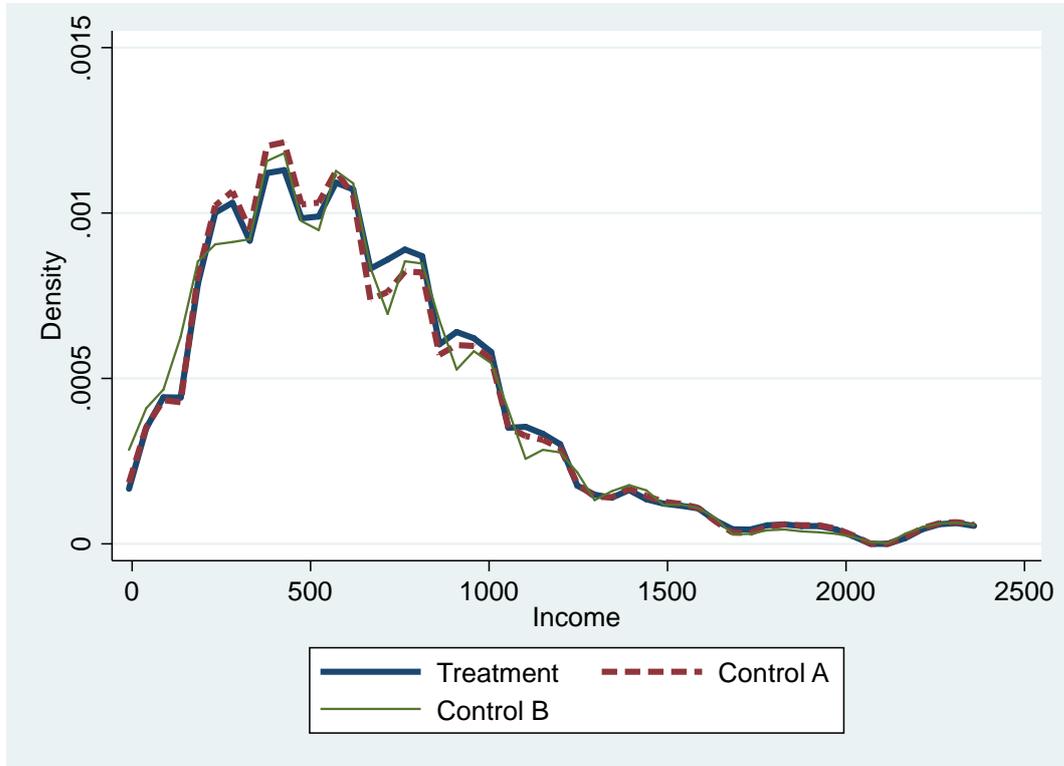

Fig.4. Kernel density for household income.

TABLE 1

BALANCE CHECK: MEAN DIFFERENCE TEST BETWEEN TREATMENT AND CONTROL GROUPS IN THE FIRST SURVEY.

| | Description | (1) Treatment | (2) Control A | (3) Control B | Absolute t-values (1) vs (2) | Absolute t-values (1) vs (3) |
|---|---|---|---|---|---|---|
| HAPPY | Level of happiness, 1(very unhappy) – 5(Very happy) | 3.17 | 3.13 | 3.18 | 1.08 | 0.27 |
| SATISF | Level of life satisfaction 1(very unsatisfied) – 5(Very satisfied) | 3.07 | 3.05 | 3.09 | 0.64 | 0.54 |
| RANK_LOCAL | Income position in the same residential prefecture (the bottom is 0.01, and the top 0.99) A larger value indicates that a subjective income position is higher than a real one. | 0.49 | 0.48 | 0.47 | 1.07 | 1.58 |
| RANK_AGE | Income position in the same cohort (the bottom is 0.01, and the top 0.99) A larger value indicates that a subjective income position is higher than a real one. | 0.48 | 0.47 | 0.47 | 1.35 | 1.75* |
| RANK_EDU | Income position in the same educational background (the bottom is 0.01, and the top 0.99) A larger value indicates that a subjective income position is higher than a real one. | 0.49 | 0.48 | 0.48 | 1.20 | 1.53 |
| Trust | Most people can be trusted. 1 (strongly disagree) – 5 (strongly agree). | 3.15 | 3.14 | 3.14 | 0.28 | 0.11 |
| Reciprocity | If someone does me a favour, I am prepared to return it 1 (strongly disagree) – 5 (strongly agree). | 4.02 | 4.02 | 4.01 | 0.52 | 0.09 |
| Male | It has 1 if respondent is male, otherwise 0. | 0.53 | 0.51 | 0.53 | 1.10 | 0.27 |
| UUNIV | It has 1 if respondent graduated from university or more, otherwise 0. | 0.24 | 0.29 | 0.24 | 0.21 | 0.14 |
| AGE | Ages | 44.9 | 44.0 | 44.5 | 1.02 | 0.93 |
| Observations | | 2,341 | 2,182 | 1,134 | | |

*Notes*: Control A is the group in which information on cross-country comparison is provided, whereas Control B is the group in which any information is provided. * indicates statistical significance at the 10% level

TABLE 2

HAPPINESS IS A DEPENDENT VARIABLE (FIXED EFFECTS MODEL)

|  | Treatment | | | Control A | | | Control B | | |
| --- | --- | --- | --- | --- | --- | --- | --- | --- | --- |
|  | (1) | (2) | (3) | (4) | (5) | (6) | (7) | (8) | (9) |
| SECOND* RANK_LOCAL | 0.175*** (0.05) | | | 0.096 (0.06) | | | −0.038 (0.07) | | |
| SECOND* RANK_AGE | | 0.143** (0.05) | | | 0.092 (0.06) | | | −0.005 (0.07) | |
| SECOND* RANK_EDU | | | 0.156*** (0.05) | | | 0.069 (0.06) | | | −0.019 (0.07) |
| SECOND | −0.103*** (0.04) | −0.086** (0.03) | −0.094*** (0.03) | −0.139*** (0.03) | −0.137*** (0.03) | −0.128*** (0.03) | −0.035 (0.04) | −0.050 (0.04) | −0.059 (0.04) |
| $R^2$ | 0.03 | 0.04 | 0.04 | 0.01 | 0.01 | 0.01 | 0.001 | 0.0002 | 0.001 |
| Group | 2,331 | 2,331 | 2,331 | 2,170 | 2,170 | 2,170 | 1,126 | 1,126 | 1,126 |
| Observations | 4,662 | 4,662 | 4,662 | 4,340 | 4,340 | 4,340 | 2,252 | 2,252 | 2,252 |

*Note*: *, **, and *** suggest statistical significance at the 10%, 5%, and 1% levels, respectively. The numbers in parentheses are the robust standard errors. Time-invariant individual characteristics are controlled by the fixed-effects model.

TABLE 3

LIFE SATISFACTION IS A DEPENDENT VARIABLE (FIXED EFFECTS MODEL)

|  | Treatment | | | Control A | | | Control B | | |
| --- | --- | --- | --- | --- | --- | --- | --- | --- | --- |
|  | (1) | (2) | (3) | (4) | (5) | (6) | (7) | (8) | (9) |
| SECOND* RANK_LOCAL | 0.146** (0.06) | | | 0.099 (0.06) | | | −0.011 (0.08) | | |
| SECOND* RANK_AGE | | 0.134** (0.06) | | | 0.077 (0.06) | | | 0.069 (0.07) | |
| SECOND* RANK_EDU | | | 0.124** (0.06) | | | 0.069 (0.06) | | | 0.014 (0.07) |
| SECOND | −0.112*** (0.03) | −0.105*** (0.03) | −0.101*** (0.03) | −0.183*** (0.04) | −0.172*** (0.04) | −0.172*** (0.04) | −0.067 (0.04) | −0.099** (0.04) | −0.071 (0.04) |
| $R^2$ | 0.04 | 0.05 | 0.03 | 0.01 | 0.01 | 0.01 | 0.001 | 0.02 | 0.002 |
| Group | 2,331 | 2,331 | 2,331 | 2,170 | 2,170 | 2,170 | 1,126 | 1,126 | 1,126 |
| Observations | 4,662 | 4,662 | 4,662 | 4,340 | 4,340 | 4,340 | 2,252 | 2,252 | 2,252 |

*Note*: *, **, and *** suggest statistical significance at the 10%, 5%, and 1% levels, respectively. The numbers in parentheses are the robust standard errors. Time-invariant individual characteristics are controlled by the fixed-effects model.

# TABLE 4

## HAPPINESS IS A DEPENDENT VARIABLE (FIXED EFFECTS MODEL) SUB-SAMPLE OF RECIPROCAL RESPONDENTS AND OTHERS.

### PANEL A

### TREATMENT

|  | Reciprocal | | | Not reciprocal | | |
|---|---|---|---|---|---|---|
|  | (1) | (2) | (3) | (4) | (5) | (6) |
| SECOND* RANK_LOCAL | 0.127 (0.10) |  |  | 0.198*** (0.06) |  |  |
| SECOND* RANK_AGE |  | 0.091 (0.09) |  |  | 0.167*** (0.06) |  |
| SECOND* RANK_EDU |  |  | 0.131 (0.10) |  |  | 0.166** (0.07) |
| SECOND | −0.069 (0.06) | −0.049 (0.06) | −0.070 (0.06) | −0.119*** (0.04) | −0.104*** (0.04) | −0.104*** (0.04) |
| Within- $R^2$ | 0.02 | 0.03 | 0.02 | 0.03 | 0.04 | 0.03 |
| Group | 760 | 760 | 760 | 1,571 | 1,571 | 1,571 |
| Observations | 1,520 | 1,520 | 1,520 | 3,142 | 3,142 | 3,142 |

### PANEL B

### GROUP A (CONTROL WITH INFORMATION OF CROSS-COUNTRIES COMPARISON)

|  | Reciprocal | | | Not reciprocal | | |
|---|---|---|---|---|---|---|
|  | (1) | (2) | (3) | (4) | (5) | (6) |
| SECOND* RANK_LOCAL | 0.121 (0.10) |  |  | 0.082 (0.07) |  |  |
| SECOND* RANK_AGE |  | 0.089 (0.10) |  |  | 0.098 (0.07) |  |
| SECOND* RANK_EDU |  |  | 0.071 (0.10) |  |  | 0.067 (0.07) |
| SECOND | −0.153** (0.06) | −0.137** (0.06) | −0.127** (0.06) | −0.132*** (0.04) | −0.137*** (0.04) | −0.128*** (0.04) |
| Within- $R^2$ | 0.01 | 0.01 | 0.01 | 0.01 | 0.01 | 0.01 |
| Group | 741 | 741 | 741 | 1,429 | 1,429 | 1,429 |
| Observations | 1,482 | 1,482 | 1,482 | 2,858 | 2,858 | 2,858 |

*Note*: *, **, and *** suggest statistical significance at the 10%, 5%, and 1% levels, respectively. The numbers in parentheses are the robust standard errors. Time-invariant individual characteristics are controlled by the fixed-effects model.

TABLE 5

LIFE SATISFACTION IS A DEPENDENT VARIABLE (FIXED EFFECTS MODEL)
SUB-SAMPLE OF RECIPROCAL RESPONDENTS AND OTHERS.

PANEL A

TREATMENT

|  | Reciprocal | | | Not reciprocal | | |
|---|---|---|---|---|---|---|
|  | (1) | (2) | (3) | (4) | (5) | (6) |
| SECOND* RANK_LOCAL | 0.139 (0.10) | | | 0.153** (0.07) | | |
| SECOND* RANK_AGE | | 0.133 (0.09) | | | 0.137** (0.07) | |
| SECOND* RANK_EDU | | | 0.106 (0.10) | | | 0.135* (0.07) |
| SECOND | −0.134** (0.06) | −0.129** (0.06) | −0.116* (0.06) | −0.102** (0.04) | −0.094** (0.04) | −0.094** (0.04) |
| Within- $R^2$ | 0.02 | 0.03 | 0.02 | 0.04 | 0.05 | 0.04 |
| Group | 760 | 760 | 760 | 1,571 | 1,571 | 1,571 |
| Observations | 1,520 | 1,520 | 1,520 | 3,142 | 3,142 | 3,142 |

PANEL B

GROUP A (CONTROL WITH INFORMATION OF CROSS-COUNTRIES COMPARISON)

|  | Reciprocal | | | Not reciprocal | | |
|---|---|---|---|---|---|---|
|  | (1) | (2) | (3) | (4) | (5) | (6) |
| SECOND* RANK_LOCAL | 0.191* (0.11) | | | 0.056 (0.07) | | |
| SECOND* RANK_AGE | | 0.135 (0.10) | | | 0.053 (0.07) | |
| SECOND* RANK_EDU | | | 0.116 (0.11) | | | 0.049 (0.07) |
| SECOND | −0.258*** (0.07) | −0.230*** (0.07) | −0.222*** (0.07) | −0.147*** (0.04) | −0.145*** (0.04) | −0.147*** (0.04) |
| Within- $R^2$ | 0.03 | 0.02 | 0.02 | 0.01 | 0.01 | 0.01 |
| Group | 741 | 741 | 741 | 1,429 | 1,429 | 1,429 |
| Observations | 1,482 | 1,482 | 1,482 | 2,858 | 2,858 | 2,858 |

*Note*: *, **, and *** suggest statistical significance at the 10%, 5%, and 1% levels, respectively. The numbers in parentheses are the robust standard errors. Time-invariant individual characteristics are controlled by the fixed-effects model.

# TABLE 6

## HAPPINESS IS A DEPENDENT VARIABLE (FIXED EFFECTS MODEL) SUB-SAMPLE OF RESPONDENTS WHO TRUST OTHERS AND THOSE WHO DO NOT TRUST OTHERS.

### PANEL A

### TREATMENT

|  | Trust | | | Not trust | | |
|---|---|---|---|---|---|---|
|  | (1) | (2) | (3) | (4) | (5) | (6) |
| SECOND* RANK_LOCAL | 0.126* (0.07) | | | 0.238*** (0.08) | | |
| SECOND* RANK_AGE | | 0.092 (0.07) | | | 0.211*** (0.08) | |
| SECOND* RANK_EDU | | | 0.103 (0.07) | | | 0.217*** (0.08) |
| SECOND | −0.111** (0.05) | −0.093** (0.04) | −0.098** (0.05) | −0.105** (0.04) | −0.089** (0.04) | −0.096** (0.04) |
| Within-$R^2$ | 0.04 | 0.04 | 0.04 | 0.02 | 0.02 | 0.02 |
| Group | 1,055 | 1,055 | 1,055 | 1,276 | 1,276 | 1,276 |
| Observations | 2,110 | 2,110 | 2,110 | 2,552 | 2,552 | 2,552 |

### PANEL B

### GROUP A (CONTROL WITH INFORMATION OF CROSS-COUNTRIES COMPARISON)

|  | Trust | | | Not trust | | |
|---|---|---|---|---|---|---|
|  | (1) | (2) | (3) | (4) | (5) | (6) |
| SECOND* RANK_LOCAL | 0.116 (0.09) | | | 0.129 (0.08) | | |
| SECOND* RANK_AGE | | 0.089 (0.08) | | | 0.154* (0.08) | |
| SECOND* RANK_EDU | | | 0.067 (0.09) | | | 0.112 (0.08) |
| SECOND | −0.206*** (0.05) | −0.192*** (0.05) | −0.180*** (0.05) | −0.106** (0.05) | −0.114*** (0.04) | −0.102** (0.05) |
| Within-$R^2$ | 0.04 | 0.04 | 0.04 | 0.02 | 0.02 | 0.02 |
| Group | 1,006 | 1,006 | 1,006 | 1,164 | 1,164 | 1,164 |
| Observations | 2,012 | 2,012 | 2,012 | 2,328 | 2,328 | 2,328 |

*Note*: *, **, and *** suggest statistical significance at the 10%, 5%, and 1% levels, respectively. The numbers in parentheses are the robust standard errors. Time-invariant individual characteristics are controlled by the fixed-effects model.

TABLE 7

LIFE SATISFACTION IS A DEPENDENT VARIABLE (FIXED EFFECTS MODEL)
SUB-SAMPLE OF RESPONDENTS WHO TRUST OTHERS AND THOSE WHO DO NOT TRUST OTHERS.

PANEL A

TREATMENT

|  | Trust | | | Not trust | | |
|---|---|---|---|---|---|---|
|  | (1) | (2) | (3) | (4) | (5) | (6) |
| SECOND* RANK_LOCAL | 0.113 (0.08) | | | 0.201** (0.08) | | |
| SECOND* RANK_AGE | | 0.112 (0.08) | | | 0.187** (0.08) | |
| SECOND* RANK_EDU | | | 0.091** (0.08) | | | 0.173** (0.08) |
| SECOND | −0.134** (0.05) | −0.134** (0.05) | −0.122** (0.05) | −0.103** (0.04) | −0.094** (0.04) | −0.092** (0.04) |
| Within- $R^2$ | 0.04 | 0.04 | 0.04 | 0.02 | 0.02 | 0.02 |
| Group | 1,055 | 1,055 | 1,055 | 1,276 | 1,276 | 1,276 |
| Observations | 2,110 | 2,110 | 2,110 | 2,552 | 2,552 | 2,552 |

PANEL B

GROUP A (CONTROL WITH INFORMATION OF CROSS-COUNTRIES COMPARISON)

|  | Trust | | | Not trust | | |
|---|---|---|---|---|---|---|
|  | (1) | (2) | (3) | (4) | (5) | (6) |
| SECOND* RANK_LOCAL | 0.163* (0.09) | | | 0.109 (0.09) | | |
| SECOND* RANK_AGE | | 0.147 (0.09) | | | 0.089 (0.08) | |
| SECOND* RANK_EDU | | | 0.119 (0.09) | | | 0.076 (0.09) |
| SECOND | −0.287*** (0.06) | −0.279*** (0.06) | −0.264*** (0.06) | −0.126*** (0.05) | −0.116** (0.05) | −0.117*** (0.05) |
| Within- $R^2$ | 0.04 | 0.04 | 0.04 | 0.02 | 0.02 | 0.02 |
| Group | 1,006 | 1,006 | 1,006 | 1,164 | 1,164 | 1,164 |
| Observations | 2,012 | 2,012 | 2,012 | 2,328 | 2,328 | 2,328 |

*Note*: *, **, and *** suggest statistical significance at the 10%, 5%, and 1% levels, respectively. The numbers in parentheses are the robust standard errors. Time-invariant individual characteristics are controlled by the fixed-effects model.

TABLE 8

HAPPINESS IS A DEPENDENT VARIABLE (FIXED EFFECTS MODEL)
SUB-SAMPLE OF RESPONDENTS WHO GRADUATED FROM UNIVERSITY, AND OTHERS.

PANEL A

TREATMENT

|  | University | | | Others | | |
|---|---|---|---|---|---|---|
|  | (1) | (2) | (3) | (4) | (5) | (6) |
| SECOND* RANK_LOCAL | 0.316*** (0.11) | | | 0.123** (0.06) | | |
| SECOND* RANK_AGE | | 0.266** (0.11) | | | 0.094 (0.06) | |
| SECOND* RANK_EDU | | | 0.289** (0.11) | | | 0.115* (0.06) |
| SECOND | −0.142** (0.07) | −0.120* (0.07) | −0.116* (0.07) | −0.089** (0.04) | −0.074** (0.03) | −0.087** (0.04) |
| Within- $R^2$ | 0.04 | 0.05 | 0.03 | 0.02 | 0.03 | 0.02 |
| Group | 551 | 551 | 551 | 1,780 | 1,780 | 1,780 |
| Observations | 1,102 | 1,102 | 1,102 | 3,560 | 3,560 | 3,560 |

PANEL B

GROUP A (CONTROL WITH INFORMATION OF CROSS-COUNTRIES COMPARISON)

|  | University | | | Others | | |
|---|---|---|---|---|---|---|
|  | (1) | (2) | (3) | (4) | (5) | (6) |
| SECOND* RANK_LOCAL | 0.193 (0.11) | | | 0.064 (0.06) | | |
| SECOND* RANK_AGE | | 0.105 (0.11) | | | 0.090 (0.07) | |
| SECOND* RANK_EDU | | | 0.098 (0.12) | | | 0.059 (0.07) |
| SECOND | −0.194*** (0.07) | −0.150** (0.07) | −0.140** (0.07) | −0.123*** (0.04) | −0.134*** (0.04) | −0.124*** (0.04) |
| Within- $R^2$ | 0.03 | 0.02 | 0.01 | 0.007 | 0.01 | 0.006 |
| Group | 515 | 515 | 515 | 1,655 | 1,655 | 1,655 |
| Observations | 1,030 | 1,030 | 1,030 | 3,310 | 3,310 | 3,310 |

*Note*: *, **, and *** suggest statistical significance at the 10%, 5%, and 1% levels, respectively. The numbers in parentheses are the robust standard errors. Time-invariant individual characteristics are controlled by the fixed-effects model.

TABLE 9

LIFE SATISFACTION IS A DEPENDENT VARIABLE (FIXED EFFECTS MODEL)
SUB-SAMPLE OF RESPONDENTS WHO GRADUATED FROM UNIVERSITY, AND OTHERS.

PANEL A

TREATMENT

|  | University | | | Others | | |
| --- | --- | --- | --- | --- | --- | --- |
|  | (1) | (2) | (3) | (4) | (5) | (6) |
| SECOND* RANK_LOCAL | 0.348*** (0.11) | | | 0.079 (0.07) | | |
| SECOND* RANK_AGE | | 0.283** (0.11) | | | 0.085 (0.07) | |
| SECOND* RANK_EDU | | | 0.251** (0.11) | | | 0.085 (0.06) |
| SECOND | −0.203*** (0.07) | −0.173* (0.07) | −0.142** (0.07) | −0.084** (0.04) | −0.085** (0.04) | −0.088** (0.04) |
| Within- $R^2$ | 0.06 | 0.08 | 0.05 | 0.02 | 0.03 | 0.02 |
| Group | 551 | 551 | 551 | 1,780 | 1,780 | 1,780 |
| Observations | 1,102 | 1,102 | 1,102 | 3,560 | 3,560 | 3,560 |

PANEL B

GROUP A (CONTROL WITH INFORMATION OF CROSS-COUNTRIES COMPARISON)

|  | University | | | Others | | |
| --- | --- | --- | --- | --- | --- | --- |
|  | (1) | (2) | (3) | (4) | (5) | (6) |
| SECOND* RANK_LOCAL | 0.079 (0.13) | | | 0.111 (0.07) | | |
| SECOND* RANK_AGE | | 0.003 (0.13) | | | 0.107 (0.07) | |
| SECOND* RANK_EDU | | | −0.018 (0.13) | | | 0.098 (0.07) |
| SECOND | −0.193** (0.08) | −0.153* (0.08) | −0.143* (0.08) | −0.182*** (0.04) | −0.179*** (0.04) | −0.181*** (0.04) |
| Within- $R^2$ | 0.01 | 0.004 | 0.003 | 0.02 | 0.02 | 0.01 |
| Group | 515 | 515 | 515 | 1,655 | 1,655 | 1,655 |
| Observations | 1,030 | 1,030 | 1,030 | 3,310 | 3,310 | 3,310 |

*Note*: *, **, and *** suggest statistical significance at the 10%, 5%, and 1% levels, respectively. The numbers in parentheses are the robust standard errors. Time-invariant individual characteristics are controlled by the fixed-effects model.

TABLE 10

HAPPINESS IS A DEPENDENT VARIABLE (FIXED EFFECTS MODEL)

PANEL A

TREATMENT

|  | Age=<45 | | | Age>45 | | |
|---|---|---|---|---|---|---|
|  | (1) | (2) | (3) | (4) | (5) | (6) |
| SECOND* RANK_LOCAL | 0.183** (0.08) |  |  | 0.168** (0.07) |  |  |
| SECOND* RANK_AGE |  | 0.161* (0.08) |  |  | 0.126* (0.07) |  |
| SECOND* RANK_EDU |  |  | 0.166* (0.08) |  |  | 0.146** (0.07) |
| SECOND | −0.107** (0.05) | −0.099** (0.05) | −0.099** (0.05) | −0.099** (0.04) | −0.074* (0.04) | −0.088** (0.04) |
| Within- $R^2$ | 0.03 | 0.03 | 0.02 | 0.03 | 0.03 | 0.03 |
| Group | 1,188 | 1,188 | 1,188 | 1,143 | 1,143 | 1,143 |
| Observations | 2,376 | 2,376 | 2,376 | 2,286 | 2,286 | 2,286 |

PANEL B

GROUP A (CONTROL WITH INFORMATION OF CROSS-COUNTRIES COMPARISON)

|  | Age=<45 | | | Age>45 | | |
|---|---|---|---|---|---|---|
|  | (1) | (2) | (3) | (4) | (5) | (6) |
| SECOND* RANK_LOCAL | 0.044 (0.09) |  |  | 0.133* (0.08) |  |  |
| SECOND* RANK_AGE |  | 0.032 (0.09) |  |  | 0.137** (0.07) |  |
| SECOND* RANK_EDU |  |  | 0.009 (0.09) |  |  | 0.107 (0.07) |
| SECOND | −0.133** (0.05) | −0.128** (0.05) | −0.120** (0.05) | −0.141*** (0.04) | −0.139*** (0.04) | −0.128*** (0.04) |
| Within- $R^2$ | 0.005 | 0.005 | 0.005 | 0.02 | 0.03 | 0.02 |
| Group | 1,097 | 1,097 | 1,097 | 1,073 | 1,073 | 1,073 |
| Observations | 2,194 | 2,194 | 2,194 | 2,146 | 2,146 | 2,146 |

*Note*: *, **, and *** suggest statistical significance at the 10%, 5%, and 1% levels, respectively. The numbers in parentheses are the robust standard errors. Time-invariant individual characteristics are controlled by the fixed-effects model.

# TABLE 11

## LIFE SATISFACTION IS A DEPENDENT VARIABLE (FIXED EFFECTS MODEL)

### PANEL A

### TREATMENT

|  | Age=<45 | | | Age>45 | | |
| --- | --- | --- | --- | --- | --- | --- |
|  | (1) | (2) | (3) | (4) | (5) | (6) |
| SECOND* RANK_LOCAL | 0.262*** (0.08) |  |  | 0.046 (0.07) |  |  |
| SECOND* RANK_AGE |  | 0.260** (0.08) |  |  | 0.024 (0.07) |  |
| SECOND* RANK_EDU |  |  | 0.238*** (0.0)) |  |  | 0.027 (0.08) |
| SECOND | −0.159*** (0.05) | −0.161*** (0.05) | −0.147*** (0.05) | −0.067 (0.05) | −0.054 (0.04) | −0.057 (0.05) |
| Within- $R^2$ | 0.04 | 0.05 | 0.04 | 0.01 | 0.006 | 0.005 |
| Group | 1,188 | 1,188 | 1,188 | 1,143 | 1,143 | 1,143 |
| Observations | 2,376 | 2,376 | 2,376 | 2,286 | 2,286 | 2,286 |

### PANEL B

### GROUP A (CONTROL WITH INFORMATION OF CROSS-COUNTRIES COMPARISON)

|  | Age=<45 | | | Age>45 | | |
| --- | --- | --- | --- | --- | --- | --- |
|  | (1) | (2) | (3) | (4) | (5) | (6) |
| SECOND* RANK_LOCAL | 0.069 (0.09) |  |  | 0.106 (0.08) |  |  |
| SECOND* RANK_AGE |  | 0.061 (0.09) |  |  | 0.084 (0.08) |  |
| SECOND* RANK_EDU |  |  | −0.002 (0.09) |  |  | 0.108 (0.08) |
| SECOND | −0.203*** (0.05) | −0.199*** (0.05) | −0.173*** (0.05) | −0.152*** (0.05) | −0.139*** (0.05) | −0.156*** (0.04) |
| Within- $R^2$ | 0.01 | 0.01 | 0.005 | 0.02 | 0.02 | 0.02 |
| Group | 1,097 | 1,097 | 1,097 | 1,073 | 1,073 | 1,073 |
| Observations | 2,194 | 2,194 | 2,194 | 2,146 | 2,146 | 2,146 |

*Note*: *, **, and *** suggest statistical significance at the 10%, 5%, and 1% levels, respectively. The numbers in parentheses are the robust standard errors. Time-invariant individual characteristics are controlled by the fixed-effects model.